\documentclass[a4paper,11pt]{article}
\usepackage[T1]{fontenc}
\usepackage[utf8]{inputenc}
\usepackage{lmodern}
\usepackage{amsmath}
\usepackage{amssymb}
\usepackage{amsthm}
\usepackage[usenames,dvipsnames]{color}
\usepackage{xcolor}
\usepackage{bm}
\usepackage{cite}
\usepackage[usenames,dvipsnames]{color}
\usepackage{fullpage}
\usepackage{bigfoot}
\date{\today}
\title{Thermodynamics of inhomogeneous imperfect quantum gases in harmonic traps}
\author{Krzysztof Myśliwy and Marek Napiórkowski \\ 
Institute of Theoretical Physics, Faculty of Physics \\ University of Warsaw, 
 Pasteura 5, 02-093 Warszawa, Poland}

\begin{document}

\def\bk{\bf k}
\def\bm{\bf m}
\newcommand{\be}{\begin{equation}}
\newcommand{\ee}{\end{equation}}
\newcommand{\beq}{\begin{eqnarray}}
\newcommand{\eeq}{\end{eqnarray}}
\maketitle

\begin{abstract}
We discuss thermodynamic properties of harmonically trapped imperfect quantum gases. The spatial inhomogeneity of these systems imposes a redefinition of the mean-field interparticle potential energy as compared to the homogeneous case. In our approach, it takes the form $\frac{a}{2} N^2 \, \omega^d$, where $N$ is the number of particles, $\omega$ - the harmonic trap frequency, $d$ - system's dimensionality, and $a$ is a parameter characterizing the interparticle interaction. We provide arguments that this model corresponds to the limiting case of a long-ranged interparticle potential of vanishingly small amplitude. This conclusion is drawn from a computation similar to the well-known Kac scaling procedure, which is presented here in a form adapted to the case of an isotropic harmonic trap.
We show that within the model, the imperfect gas of trapped repulsive bosons undergoes the Bose-Einstein condensation provided $d>1$.
The main result of our analysis is that in $d=1$ the gas of attractive imperfect fermions with $a=-a_{F}<0$ is thermodynamically equivalent to the gas of repulsive bosons with $a=a_{B}>0$ provided the parameters $a_{F}$ and $a_{B}$ fulfill the relation $a_{B}+a_{F}=\hbar$.  
This result supplements similar recent conclusion about thermodynamic equivalence of two-dimensional uniform imperfect repulsive 
Bose and attractive Fermi gases.  
\end{abstract}
\newpage 
\section{Introduction}

The properties of trapped quantum gases, bosons or fermions,  have been the subject of intense experimental and theoretical research in recent years. It brought a wealth of important results and the volume of the relevant literature evades any attempts to provide the reader with a selection of reasonable size. We would nevertheless like point out to Refs. [1-23] 
as an example of a representative (yet still insufficient) sample on the subject. \\ \indent From the point of view of thermodynamics, it is the inclusion of external fields that makes the theoretical description of these phenomena quite distinctive. This is due to the inherent inhomogeneity of trapped systems. Understanding how inhomogeneities manifest themselves on the level of a macroscopic (in particular, thermodynamic) description is of fundamental relevance for numerous fields, not only physics of cold gases. Motivated by this general challenge, we analyze and discuss a soluble model of interacting quantum particles, adapted to the case where the particles are placed in an external field; for the sake of simplicity, we restrict ourselves to the well-studied case of a harmonic trap. "Quantum" means here predominantly the explicit inclusion of quantum statistics, either Bose-Einstein or Fermi-Dirac, in the analysis.
\subsection{The homogeneous imperfect gas} \indent Even when inhomogeneity is absent, capturing the essence of the phenomena related to the realm of quantum condensed matter is not easy. This is why discussing exactly soluble, simple models is of a great value. One of these is provided by the so-called imperfect quantum gas model. 

The model was proposed several years ago and is well established in the literature [24-32]. 
In this description, the total potential energy of interparticle interactions is assumed to take the mean-field form and the system's Hamiltonian is 
\be
\label{HPG}
H_{imp} = \sum_{\bk} \frac{\hbar^2 \bk^{2}}{2m} n_{\bk}  + \frac{a}{2} \frac{N^2}{V}   \quad, 
\ee 
where $N = \sum\limits_{\bk}n_{\bk}$, $n_{\bk}$ is the occupation number of one-particle state with momentum $\hbar{\bf{k}}$. The parameter $a$ measures the strength of two-particle interaction and serves as the proportionality coefficient in the mean-field expression for the total potential energy $a N^2/2V$. The sign of 
$a$ indicates the nature of the interparticle interaction, i.e. the interaction is attractive for $a<0$ and repulsive for $a>0$. Thus, in the case of imperfect, repulsive bosons we have $a=a_{B}>0$ and $n_{\bk}\,=\,0,1, \dots \infty$, while for the imperfect attractive fermions  $a=-a_{F}<0$ and $n_{\bk}\,=\,0,1$. One can then show \cite{NP2017} that the thermodynamics of the imperfect attractive spinless fermions exists only for 
$d \leq 2$ and that exactly at $d=2$ the thermodynamics exists only for negative values of chemical potential: $\mu <0$. No such restrictions are present in the case of imperfect, repulsive bosons. Remarkably, the thermodynamics of two-dimensional attractive fermions is identical to the thermodynamics of two-dimensional repulsive bosons provided the corresponding parameters $a_{F}$ and $a_{B}$ 
fulfil the relation \cite{NP2017}
\be
\label{symhom}
a_{F} + a_{B} = a_{0} = \frac{h^2}{2\pi m} \quad. 
\ee
It follows from the above relation that the thermodynamics of two-dimensional imperfect attractive fermions with $a_{F}=a_{0}$ is identical to the thermodynamics of two-dimensional {\it perfect} Bose gas. And, symmetrically, the thermodynamics of two-dimensional imperfect repulsive bosons  with $a_{B}=a_{0}$ is identical to the thermodynamics of two-dimensional {\it perfect} Fermi gas. This equivalence can be intuitively understood as the result of mutual balance between the exchange (statistical) interaction and the physical interaction.  
\subsection{Outline}
The above thermodynamic equivalence holds for {\it uniform} imperfect quantum gases in $d=2$. In this article we want to address the question whether analogous results can hold for {\it nonuniform} imperfect quantum gases enclosed in harmonic traps. This issue is of practical relevance since many of the experimental results discussed in the literature refer to trapped quantum gases. From the theoretical side, it is interesting to investigate how the external potential influences the aforementioned balance between the exchange  and the \emph{physical} interactions, and how the role of dimensionality comes into play in this inhomogeneous setup, also with respect to stability. The approach we adopt is based on the postulate that the total interaction energy term has similar formal properties to that of the homogeneous imperfect gas. Physically, these properties can be attributed to the Kac limiting procedure \cite{KUH1,HL1976}, which corresponds to the case of an extremely weakly interacting gas with a long-ranged interaction potential. The adaptation we give was proposed independently in \cite{Zagreb}. 
To complement the discussion, we also investigate the emergence of the Bose-Einstein condensation in an imperfect gas of trapped, repulsive bosons, analogously to the homogeneous case \cite{NJN1}; for a thorough discussion, see \cite{Zagreb}. 
\\\indent Our work has the following structure. In Sec. 2 we introduce the aforementioned modification of the imperfect quantum gas Hamiltonian which applies to the case when the particles are located in an isotropic harmonic trap. We present a method of evaluating the grand canonical partition function for this model, discuss the system's stability, and derive the formulae for the relevant thermodynamic quantities. In Subsec. 2.2 we briefly discuss the phenomenon of the Bose-Einstein condensation in a system of trapped, repulsive bosons, observed when $d>1$. In Subsec. 2.3 we show that within this approach, the trapped fermionic particles interacting via an attractive potential are indeed thermodynamically equivalent to the system of repulsively interacting trapped bosons, provided the dimensionality of the systems is $d=1$ and the coupling constants are appropriately adjusted. Sec. 3 is devoted to a short summary and discussion. The Appendix presents our adaptation of the Kac scaling to the case of the harmonic trap. The postulated form of the interaction energy emerges there in a straightforward way.

\section{Imperfect gas in a harmonic trap}
\subsection{Adaptation of the imperfect gas model to harmonic traps}
We aim at evaluating the thermodynamic properties of the $d$-dimensional imperfect gas of $N$ interacting spinless quantum particles of mass $m$ 
(either fermions of bosons) subject to the external harmonic potential 
\begin{equation}
  W(r)=\frac{m \omega^2 r^2 }{2} \quad,
\end{equation} 
 where $r$ is the radial coordinate in $\mathbb{R}^d$ and $\omega$ - the trapping frequency. In the absence of interparticle interactions this model yields the ideal trapped gas \cite{Posa2006}.  When the interparticle interactions are to be included in the spirit to the imperfect quantum gas model, the imperfect Hamiltonian in Eq.(\ref{HPG}) ceases to be directly applicable  as the available volume for the trapped particles is considered infinite from the very beginning. On the other hand, we wish to keep the $N^2$ factor in the mean-field expression for the gas potential energy because its proportionality to the number of pairs of interacting particles reflects the physical meaning of this approach \cite{Zagreb,Huang1,Davies1,BP1983,Zag2001}. The potential energy has to give a nontrivial contribution to the intensive thermodynamic quantities of the trapped gas in the appropriately defined limit  which plays the role of the standard thermodynamic limit. For harmonically trapped gases this limit corresponds to $\omega \rightarrow 0, N \rightarrow \infty$ with $N \omega^d =\rho_{\omega}$ fixed \cite{RRB2004,RR2005,Posa2006,Zagreb,ASY2018}, and is also called the thermodynamic limit. A generalization of \eqref{HPG} in the spirit discussed above is relatively clear in $d=1$, which also turns out to be of special relevance in our analysis. The harmonic trap offers the energy scale $\hbar \omega$, analogously to the particle-in-a-box energy scale  $a_0/V^{\frac{2}{d}}$. Thus in $d=1$ the system's interaction energy is postulated in the simple form 
 \begin{equation}
 \label{imp1} 
 H_{int} = \tilde{a} \,\hbar \omega \,\frac{N^2}{2} \quad. 
 \end{equation} 
 The $N^2/2$ term gives the number of pairs of interacting particles in the trap, and the dimensionless constant $\tilde{a}$ represents the ratio of the \emph{average} interaction energy of a pair of trapped particles to the energy scale provided by the trap. Our analysis is devoted to this particular implementation  of the mean-field approach to trapped systems. To provide our analysis with a broader perspective we generalize \eqref{imp1} to  arbitrary dimension in the form

\be
\label{imp}
  H_{int} = a \,\omega^d\, \frac{N^2}{2}.
\ee 
This model has been rigorously studied in the bosonic case in \cite{Zagreb}. The obvious interpretation of going from \eqref{imp1} to \eqref{imp} is that one rescales the interaction parameter $\tilde{a} \rightarrow a\, \omega^{d-1}$ which - in the limit $\omega \rightarrow 0$ - corresponds to the case of a very weak interaction. Nevertheless,  in the thermodynamic limit this interaction term remains a source of  non-trivial contribution to the thermodynamics of such systems. On the other hand, the $N^2$ scaling of the total interaction energy in (\ref{imp}) suggests a long-ranged interparticle potential as its source. This suggestion finds its support within the Kac model \cite{KUH1,HL1976}, see Appendix for details. Results obtained on the basis of (\ref{imp}) for the low-temperature equation of state of $d>1$ bosons do not reproduce some experimental and theoretical results known from the Gross -- Pitaevskii theory of weakly interacting, dilute trapped Bose gases with short-ranged interparticle potentials \cite{4x}. This discrepancy seems to indicate that the form of the interaction in (\ref{imp})  could only be considered as an approximation for weakly interacting gases at relatively high densities, resp. interacting via long-ranged potentials. We refer the discussion at the end of Sec. 2.5 for further details.
\\ \indent We emphasize that the above postulated form of the interaction is different from the one that is usually called a "mean field" in the context of inhomogeneous gases within the so-called local density approximation. There, the interaction term is chosen such that the energy density is proportional to $n^2(r)$, where $n(r)$ is the local density. The system is then locally approximated by a gas subject to an equation of state equal to that of the homogeneous imperfect gas as described in Eq. (1). This, however, is not in accord with the model we want to propose, which requires that the energy be proportional to the total number of pairs of particles in the system (\emph{global} mean field). This, as we shall see, will lead to distinct results as compared to the ones obtained by the \emph{local} version of the mean field theory. A similar distinction is encountered in \cite{SmedtZagr}.  
\\ 
\indent
 With (\ref{imp}) as the interaction term, the energy levels of the imperfect gas of $N$ particles in a harmonic trap take the following form 
\begin{equation}
\label{enlev}
  E_N(\{n_{\bm}\})=\hbar \omega \sum_{\bm} \left( \sum_{i=1}^d m_i +\frac{d}{2}\right) n_{\bm}+\frac{a}{2} \,\omega^d \,N^2 \quad, 
\end{equation} 
where $\bm$=$(m_1,\cdots, m_d)$ and the quantum numbers   $m_i=0,1,\cdots ,\infty$ correspond to the energy levels of a one-dimensional harmonic oscillator. The occupation numbers $n_{\bm}$ are constrained by  $N=\sum\limits_{\bm} n_{\bm}$; for fermions $n_{\bm}=0,1$  while for bosons $n_{\bm}=0,1,\cdots, \infty$. Correspondingly, the parameter $a=-a_{F}<0$ for attractive fermions, and $a=a_{B}>0$ for repulsive bosons. \\ \indent

\subsection{Solution method}
To analyze the thermodynamic properties of trapped, imperfect quantum gases we employ the grand canonical ensemble. The thermodynamic 
state of the system is specified by the temperature $T$ and chemical potential $\mu$. The harmonic trap frequency $\omega$ is an additional parameter with $\omega^{-d}$  playing the role of volume $V$.  In order to find the thermodynamics of this system we compute the grand canonical partition 
function 
\begin{equation}
\label{gcp}
  \Xi(T,\mu,\omega)=\sum_{N=0}^{\infty}e^{\beta\mu N} {\sum\limits_{\lbrace n_{\bm}\rbrace }}{'} e^{-\beta E_N(\{n_{\bm}\})},  
\end{equation} 
where ${\sum\limits_{\lbrace n_{\bm}\rbrace }}{'}$ denotes the sum over all values of $n_{\bm}$ satisfying the constraint 
$\sum\limits_{\bm} n_{\bm}=N$. The evaluation of (\ref{gcp}) follows the same lines as in \cite{NP2017} and makes use of the Hubbard-Stratonovich transformation \cite{Hubbard1} based on identities 
\begin{eqnarray}
\label{hubbard1}
   e^{-\alpha N^2/2} = \frac{1}{\sqrt{2\pi \alpha}}\,\int\limits_{-\infty}^{\infty} e^{-\frac{x^2}{2\alpha}-\mathrm{i}Nx} \,dx \quad,
   \eeq
   \beq
   \label{hubbard2}
   e^{\alpha N^2/2}  = \frac{1}{\sqrt{2\pi \alpha}}\,\int\limits_{-\infty}^{\infty} e^{-\frac{x^2}{2\alpha}-Nx} \,dx \quad, 
\end{eqnarray} 
which hold for $\alpha>0$. The first of these identities is suitable for repulsive bosons with $a=a_{B}>0$ while the second for attractive fermions with $a=-a_{F}<0$,  see Eqs (\ref{enlev}) and (\ref{gcp}). After inserting (\ref{hubbard1}), (\ref{hubbard2}) and (\ref{enlev}) into Eq.(\ref{gcp}) and changing the integration variables one  obtains  the following integral expressions for the grand canonical partition functions for imperfect, repulsive trapped bosons  $\Xi_{+}(T,\mu,\omega)$ and imperfect, attractive trapped fermions $\Xi_{-}(T,\mu,\omega)$
\begin{equation}
\label{int1}
  \Xi_{\zeta}(T,\mu,\omega) = (-i)^{\frac{1+\zeta}{2}} \,\sqrt{\frac{\omega^{-d}}{2\pi\beta a_{\zeta}}}\,\int\limits_{\gamma_{\zeta}} \mathrm{d}\eta \, e^{\,\omega^{-d}\,\Phi_{\zeta}(\eta;\beta,\mu)} \quad,
\end{equation}  
where $\zeta=\pm$, $\gamma_{-}$ denotes the real axis, and $\gamma_{+}$ is a contour in the complex plane, parallel to the imaginary 
axis chosen such that $\mathrm{Re}(\eta)<0$.  The quantity $a_{\zeta}$ is chosen such that $a_{+}=a_{B}$ corresponds to repulsive bosons while $a_{-}=a_{F}$ corresponds to  attractive fermions. The function $\Phi_{\zeta}(\eta;\beta,\mu)$ in 
Eq.(\ref{int1}) has the following form 
\beq
\label{psi1}
  \Phi_{\zeta}(\eta;\beta,\mu) = \zeta \left(\frac{k_B T}{\hbar}\right)^d g_{d+1}(\zeta\, e^{\eta})+\zeta \,\frac{(\eta-\beta\bar{\mu})^2}{2\beta a_{\zeta}} - \frac{(1+\zeta)}{2\, \omega^{-d}}\log(1-e^{\eta}) ,
\eeq
where $\bar{\mu}= \mu - \hbar \omega \frac{d}{2}$, and $g_{\kappa}(z)=\sum\limits_{n=1}^{\infty}\,\frac{z^n}{n^\kappa}$ is the Bose function. Note that the $\log$ term in Eq.(\ref{psi1}) is relevant only for the case of repulsive bosons in $d>1$ and accounts for the phenomenon of the Bose-Einstein condensation, see  \cite{NP1,NJN1}.
With the grand canonical partition function given by Eqs (\ref{int1}) and (\ref{psi1}) one can discuss the issue of the existence of  the thermodynamic limit  of the grand canonical free energy and its form using the method of steepest decent.

\subsection{Existence of thermodynamics of trapped imperfect gases}

In order to address the question of convergence of the integrals defining the grand canonical partition function in Eq.(\ref{int1}) and the existence of thermodynamics, we recall that for 
$\mathrm{Re}(\eta)\rightarrow -\infty$ one has  $g_{d}(\zeta e^{\eta})\rightarrow \zeta \, e^\eta$, irrespectively of $d$, and for  
$\eta\rightarrow +\infty$, $g_{\kappa}(-e^{\eta})\sim -\eta^{\kappa}$. Thus the relevant integral exists for repulsive trapped bosons.  
For fermions, on the other hand, the  convergence of the integral in Eq.(\ref{int1}) depends crucially on the large $\eta$ behaviour of $\Phi_{-}(\eta;\beta,\mu)$ in Eq.(\ref{psi1}). For $\eta$ approaching $-\infty$ one has $\Phi_{-}(\eta;\beta,\mu) \sim -\,\eta^2$ for all $d$. For $\eta\rightarrow +\infty$ one has $\Phi_{-}(\eta;\beta,\mu) \sim -\eta^2 + c \,\eta^{d+1}$ with positive (temperature dependent) constant $c$. Hence the integral converges for $d<1$ and diverges for $d>1$. If $d=1$, a closer analysis shows that $2 \beta a_{F} \,\Phi_{-}(\eta;\beta,\mu) \, = \, \frac{a_{F} - \hbar}{\hbar}\eta^2 + 2\frac {\bar{\mu}}{k_{B}T} \eta \, + \, 
\left(\frac{\bar{\mu}}{k_{B}T}\right)^2  $. Hence the integral converges if $0\leq a_{F} <\hbar$. If $a_{F}=\hbar$ one has to impose the condition $\bar{\mu}<0 $ in order to provide the convergence of the integral and thus the existence of thermodynamics. 

\subsection{Thermodynamic quantities of trapped imperfect gases}

In the limit $\omega \rightarrow 0$  the grand canonical free energy density $\Omega(T,\mu,\omega)\,\omega^d$ and the average particle density $\langle N \rangle \,\omega^d$ can be calculated for imperfect trapped attractive fermions and repulsive bosons  by evaluating the integrals in Eq. (\ref{int1})  via the method of steepest descent.  In agreement with the remarks in Subsec. 2.1 concerning the existence of thermodynamic description we restrict our considerations to the cases $d \leq 1$  when discussing fermions \cite{GBL2013}. In the following we use the short hand notation: the fermionic (bosonic) quantities are denoted  with superscript F (B) and correspond to parameter $\zeta = -(+)$. At the same time the interaction parameter $a$ is replaced by $a_{\zeta}$ with  $a_{-}=a_{F}$ or $a_{+}=a_{B}$. \\
The equation  $\Phi_{\zeta}'(\eta;\beta,\mu))|_{\eta=q}\,=\,0$ determining the saddle point $\eta=q_{\zeta}(T,\mu)$ takes the  form 
\be
\label{saddle}
  q_{\zeta} = \beta \bar{\mu} \,-\, \frac{a_{\zeta}}{\hbar^d \,\beta^{d-1}} \, g_d (\zeta e^{q_{\zeta}}) -
  \frac{\beta a_{\zeta} \zeta (1+\zeta)}{2 \omega^{-d}} \,\frac{e^{q_{\zeta}}}{1-e^{q_{\zeta}}} \quad. 
\ee
  This results in the following expressions for the harmonic density 
 \begin{equation}\label{dens}
  \rho^{\zeta}_{\omega}(T,\mu)=\langle N \rangle \omega^d = \zeta \, \left(\frac{k_B T}{\hbar}\right)^d \, g_d (\zeta e^{q_{\zeta}(T,\mu)}) \, 
  + \, \frac{1+\zeta}{2\,\omega^{-d}} \, \frac{e^{q_{\zeta}(T,\mu)}}{1-e^{q_{\zeta}(T,\mu)}}
\end{equation}  
and the free energy density
  \beq 
  \label{density}
  \beta \Omega^{\zeta}(T,\mu,\omega) \omega^d \,= -\,\zeta \,\left(\frac{k_B T}{\hbar}\right)^d  g_{d+1}(\zeta e^{q_{\zeta}(T,\mu)}) - \zeta \,\frac{a_{\zeta} \beta (\rho_{\omega}^\zeta)^2}{2}\quad.
\eeq
  
We note that the above formulae have similar structure as in the case of homogeneous imperfect gases, see \cite{NP2017}. This suggests the following correspondence when going from the trapped to the free case:  $d\rightarrow 2d$  and $\frac{\hbar} {k_B T} \rightarrow \lambda.$
The thermodynamic entropy, on the other hand, takes the form
\begin{equation}
S_{\zeta}(T,\mu,\omega)=S^{id}(T,\mu-a_{\zeta} \rho_{\omega}(T,\mu),\omega),
\end{equation} where $S^{id}$ denotes the corresponding expression for the ideal gas. In other words, the entropy as a function of $\rho_{\omega},T$ is given by 
\begin{equation}
S_{\zeta}(T,\rho_{\omega},\omega)=S^{id}(T,q(T,\rho_{\omega}),\omega)
\end{equation} which, given \eqref{dens}, does \emph{not} depend on the interaction parameter $a_{\zeta}$. This has a very simple interpretation, as the entropy is uniquely determined by the probability distribution of the occupation numbers of the one-particle orbitals. 	An analogous phenomenon is encountered in the well-known van der Waals theory of interacting attractive gases with the equation of state  $p=\frac{n k_B T}{1-n b}-\frac{a n^2}{2}$: also here the entropy per particle $s(T,n)$ depends only on $b$ and not on $a$. On the other hand, its microcanonical form is given by $S(E,V,N;a,b)=S_{id}(E+aN^2/2V,V-Nb,N)$, where $S_{id}(E,V,N)$ is the entropy of the classical ideal gas. The term $E+aN^2/2V$ can be, for the attractive gas, interpreted as the kinetic energy, and it hence stems entirely from the probability distribution of the momenta. Consequently, the contribution to the entropy of the van der Waals gas which emerges from the spatial distribution of the particles is, in fact, dependent only on $b$. The fact that the imperfect gas model admits a similar feature is responsible for the model's various formal properties. 
\subsection{Bose-Einstein condensation of trapped repulsive bosons in $d>1$}

To show the emergence of BEC in this system (for comparison, see \cite{Zagreb,NJN1}), let us note that for $d>1$ the function  $g_d(e^{q_+})$ is bounded. On the other hand, if $q_+(T,\mu)$ approaches zero, the ground state harmonic density $\rho^{0}_{\omega}=\frac{\omega^d e^{q_+}}{1-e^{q_+}}$ may account for the condensate contribution to the total harmonic density $\rho^{B}_{\omega}(T,\mu)$ when $\omega\rightarrow 0$. Recall that the bosonic quantities are defined only for $q_{+}(T,\mu) \leq 0$ which follows from Eq.(\ref{saddle}). If $q_+(T,\mu)<0$ and $\mu< \mu_{c}(T) = \frac{a_{B} \zeta(d))}{(\beta\hbar)^d}$, where $\zeta(d)$ is the Riemann zeta function,  the ground state occupation vanishes in the thermodynamic limit. If, on the other hand,  $\mu>\mu_{c}(T)$ we have $q_+(T,\mu)=0$ and one needs to include the ground state contribution 
\be \label{gcc}
\rho_{\omega}^0(T,\mu)=\frac{\mu-\mu_{c}(T)}{a_{B}}
\ee 
to the saddle point equation, and to the total harmonic density. The condition $\mu>\mu_{c}(T)$ cannot be fulfilled unless $d>1$. Thus, for $d>1$, the equation $\mu=\mu_{c}(T)$ defines the critical line in the $(T,\mu)$ space. Above this line one observes the low-temperature phase hosting the condensate. The effect of the interactions is that, contrary to the ideal gas case, the portion of the $(T,\mu)$ space with positive chemical potential becomes available for the system, curing the pathology encountered in the Bose gas under the absence of interparticle interactions. Within the imperfect gas model the effect of the external potential displays itself as the $d\rightarrow 2d$ correspondence with the homogeneous case, see remarks at the end of Subsec. 2.2. This correspondence enables the system to display the Bose-Einstein condensation in $d>1$ in accordance with the Mermin--Wagner theorem \cite{MW1966,Hoh1967,Mermin1967} which prevents continuous symmetry breaking in \emph{homogeneous} systems with dimensionalities $d\leq2$.  In fact, Bose gases subject to external fields and different boundary conditions may display BEC for various dimensionalities, including $d=1$. This is one of the manifestations of how inhomogeneities can influence a fundamental, macroscopic property of a thermodynamic system. \\ 

The formula (\ref{gcc}) reveals an important characteristic of the model we are considering. Namely, it displays a linear relation between the chemical potential and the particle number at $T=0$, $N\sim \mu$  \emph{for all $d$}. This conclusion remains in contrast with the results following from the Thomas-Fermi approximation to the Gross-Pitaevskii theory of interacting Bose gases, which predicts $N\sim \mu^{(d+2)/2}$, so that the exponent depends on $d$ \cite{4x}. This discrepancy, apart from the interpretation of \eqref{imp} as a theory of weakly interacting gases at relatively high densities, may be attributed to the absolute lack of spatial correlations induced by interactions in the imperfect gas model. Indeed, the interaction term (\ref{imp}) influences neither the entropy per particle 
$s(T,\rho_{\omega})$ nor the spatial form of ground state profile at $T=0$. The latter is given simply by a product state (which is again specific to mean-field theories). On the contrary, the spatial form of the minimizer of the Gross-Pitaevskii functional (both in the general case  and in the TF limit) changes significantly with the GP interaction parameter $g$ ($g$ being an analogous, yet in principle a different quantity than \emph{our} $a$), which hence has an impact on the correlations in the ground state, as should be expected from a genuine interaction. The failure of the imperfect gas model in reproducing spatial correlations due to interactions can be traced back to the nature of the Kac scaling which gives rise to it, see Appendix. 
\subsection{Thermodynamic equivalence of attractive fermions and repulsive bosons in $d=1$} 
In this section we explore the case $d=1$, where the function $q$ depends on $T$ and $\bar{\mu}$ only via the product $\beta \bar{\mu}$, see Eq.(\ref{saddle}). Recall also that $g_1(e^z)=-\log(1-e^z).$ For attractive imperfect fermions one then has  
\begin{equation} 
\label{fer}
q_{-} - \beta\bar{\mu} =\frac{a_{F}}{\hbar}\,\log(1+e^{q_{-}})
\end{equation} 
while for repulsive bosons
\begin{equation}
\label{bos}
q_{+} - \beta\bar{\mu} =\frac{a_{B}}{\hbar}\,\log(1-e^{q_{+}}) \quad. 
\end{equation} 
To show the equivalence we introduce the auxiliary variable  $x$ such that 
\begin{equation}
\label{fund}
  1+e^{q_{-}} = \frac{1}{1-e^x} \quad,
\end{equation} 
or equivalently 
\begin{equation} \label{x}
  x-\beta\bar{\mu} = -\left(\frac{a_{F}}{\hbar}-1\right)\,\log(1-e^x) \quad.
\end{equation} 
From the uniqueness of solutions of Eqs (\ref{fer}), (\ref{bos}), and (\ref{fund}) it follows that if 
\be
\label{podst}
a_{F}+ a_{B} = \hbar
\ee
then $x(T,\mu) = q_{+}(T,\mu)$. Moreover, it follows from Eq.(\ref{fund}) that $\log(1+e^{q_{-}})= - \log(1-e^x)$ with the consequence that if Eq.(\ref{podst}) is fulfilled then from (\ref{dens})
\begin{equation}
\label{equidens}
  \rho_{\omega}^F(T,\mu; a_F)=\rho^B_{\omega}(T,\mu; \hbar-a_F) \quad.
\end{equation} 
For clarity,  Eq.(\ref{equidens}) explicitly displays the parametric dependence of $\rho_{\omega}^{\zeta}$ on $a_{\zeta}$. Thus a trapped imperfect attractive Fermi gas with the coupling constant $a_{F}$ has the same harmonic density as the repulsive Bose gas with the  coupling constant $a_{B}=\hbar - a_{F}$. Note the boundary case $a_{F}=\hbar$ corresponds to the situation in which the trapped attractive Fermi gas has exactly the same harmonic density as the ideal Bose gas in the harmonic trap. The above correspondence holds also for the thermodynamic limit of the grand canonical free energy densities 
\begin{equation}
\label{Omega1}
  \lim\limits_{\omega \rightarrow 0} \Omega^F(T,\mu,\omega; a_F) \,\omega =  \lim\limits_{\omega \rightarrow 0} \Omega^B(T,\mu,\omega;\hbar-a_F) \,\omega. 
\end{equation}
To see this, we recall the Landen's identity \cite{GM1997} 
\begin{equation}
  g_2\left(\frac{1}{1+e^{-z}}\right)+g_2(-e^z)=-\frac{1}{2}\log^2(1+e^z) \quad 
\end{equation} 
and again represent $q_{-}(T,\mu)$ in Eq.(\ref{density}) by the auxiliary variable $x$, see Eq.(\ref{fund}), which was shown to be equal to  
$ q_{+}(T,\mu)$ if the relation (\ref{podst}) is fulfilled. In particular, (\ref{fund}) implies $\log^2(1+e^{q_{-}})=\log^2(1-e^x)$, $e^x=\frac{1}{1+e^{-q_{-}}}$. Inserting these into the Landen's identity one gets 
\begin{equation}
\label{omegaid}
    g_2\left(e^x\right)+\frac{\hbar-a}{2 \hbar}\log^2(1-e^{x})=-g_2(-e^{q_{-}})-\frac{a}{2\hbar}\log^2(1+e^{q_{-}}).
\end{equation} 
By (\ref{density}) and (\ref{x}), one finds that this is the desired result (Eq.(\ref{Omega1})) if (\ref{podst}) is satisfied. Thus, similarly to the two-dimensional homogeneous case, we demonstrated the thermodynamic equivalence of one-dimensional trapped imperfect attractive fermions and repulsive bosons.  
\section{Summary}
We have introduced a generalization of the imperfect gas model, suitable for the case of interacting particles enclosed in a harmonic trap. The main constraints imposed on this generalization were to maintain the  proportionality of the interparticle interaction potential energy to the number of pairs of particles, and to provide the existence of appropriately defined thermodynamic limit. The presented approach is meant to reflect the mean-field-like picture of reality. It is subject to an exact analytic treatment within the framework of equilibrium statistical mechanics. \\ 
\indent Having formulated the model of trapped, imperfect gases, we have derived a noteworthy result in the theory of imperfect gases, namely the thermodynamic equivalence of attractive fermions and repulsive bosons. Contrary to the free case, where this remarkable fact is observed in $d=2$, in the case of harmonically trapped imperfect gases the equivalence emerges in \emph{one-dimensional} systems. The dimensionality $d=1$ is in this case exceptional since it is the boundary dimensionality above which a system of attractively interacting imperfect trapped fermions loses its thermodynamic stability. Moreover, noting that the results for the trapped imperfect gases in dimensionality $d$ have  a similar mathematical structure as in the case of a free imperfect gases in $d'=2d$, we have discussed how does the trapped repulsive bosons undergo the Bose-Einstein condensation if the chemical potential attains a sufficiently large and positive value at a given temperature. The BEC appears only if $d>1$ which again exposes the role of this particular dimensionality. The $d\rightarrow2d$ correspondence is rooted in the presence of the external potential imposing the inhomogeneity of the studied gases, and not in the interparticle interaction. It provides an example of how the inhomogeneity of a thermodynamic system influences its fundamental properties.  \\

\noindent {\bf Acknowledgement} \\
\noindent The authors thank Professor Jaros{\l}aw Piasecki and Niels P. Benedikter for helpful discussions. M.N. acknowledges the support from National Science Center, Poland 
via grant 2014/15/B/ST3/02212. 
\\

\section{The Appendix}

Here we provide arguments that the expression for the potential energy of an imperfect trapped gas, Eq.(\ref{imp}), can be obtained via a procedure analogous to the Kac limit \cite{KUH1, HL1976} applied to the trapped quantum gas. Accordingly, one assumes that the two-body interparticle potential $v(x)$ has the form $v(x)=\gamma^d v_0 \phi(\gamma x)$, where $\gamma^{-1}$ has dimensions of length and represents the range of interaction, the amplitude $v_0$  has dimension energy$\times$volume, and $\phi$ is an integrable function with unit integral taken over infinite volume. The Kac limit $\gamma\rightarrow 0$ corresponds to interparticle potential which is both extremely weak and  long-ranged. In the homogenous case, the imperfect Hamiltonian in Eq.(\ref{HPG}) can be obtained from the second-quantized form of the Hamilton operator 
\begin{equation}
\label{secq1}
\hat{H}=\sum_{k,l}\langle k |T| l\rangle \hat{a}^{\dagger}_k \hat{a}_l+\frac{1}{2}\sum_{k,l,k',l'}\langle k, l |v|k',l'\rangle \hat{a}^{\dagger}_k \hat{a}^{\dagger}_l\hat{a}_{l'}\hat{a}_{k'} 
\end{equation}
in the Kac limit with the results result $a=v_{0}$. In the case of trapped quantum gas one replaces in Eq.(\ref{secq1})  the plane waves with the harmonic oscillator eigenstates and uses the harmonic oscillator eigenvalues instead of the kinetic energy. The imperfect Hamiltonian emerges then as the lowest order term in the $\gamma$-expansion of Eq.(\ref{secq1}). We illustrate this procedure on the one-dimensional example of exponential potential $\phi(x) = e^{-|x|}$. 

Let $\ell=\sqrt{\frac{\hbar}{m\omega}}$ denote the width of the harmonic oscillator ground state.
In what follows we assume that $\gamma^{-1}\gg \ell$ (the range of the interaction is much larger than the width of the ground state) and $\gamma \sim \omega$ (the range of the potential is proportional to the size of the thermal cloud of trapped particles). The two-body matrix element of $v$ is evaluated in the basis of harmonic oscillator eigenstates 
\beq
\label{a1}
\langle m_1, m_2 | v | m_3 ,m_4 \rangle = \nonumber \\ \gamma \, v_{0} \,\iint \mathrm{d}\xi \mathrm{d}\eta \,f_{m1,m3}(\xi)\,f_{m2,m4}(\eta)\,e^{-\xi^2-\eta^2} \phi(\ell (|\xi-\eta|)\gamma ) \quad,
\eeq
where $\xi=x/\ell$ and $\eta=y/\ell$ are dimensionless coordinates, and, by orthonormality, $\int d\xi f_{m,m'}(\xi)e^{-\xi^2}=\delta_{m,m'}$. After change of integration variables $R=\xi+\eta$, $r=\xi-\eta$ one obtains 
\beq
\langle m_1, m_2 | v | m_3 ,m_4 \rangle  =  \\                        
\frac{\gamma v_0 }{2} \,e^{\gamma^2 l^2/2} \iint \mathrm{d}R\mathrm{d}r f_{m_1,m_3}\left(\frac{R+r}{2}\right)f_{m_2,m_4}\left(\frac{R-r}{2}\right)e^{-R^2/2-(|r|-\gamma l)^2/2} \quad. \nonumber
\eeq
In the lowest order in $\gamma$ one gets 
\begin{equation}
\langle m_1 ,m_2 | v |  m_3, m_4 \rangle \approx \gamma \,v_0 \, \delta_{m_1,m_3}\,\delta_{m_2,m_4} 
\end{equation} 
which, after inserting  into the second-quantized Hamiltonian and assuming $\gamma/\omega=const$, leads to the imperfect trapped gas Hamiltonian. Note that the condition $\gamma\sim \omega$ can be imposed at the level of the saddle point expansion, which is in our case equivalent to the thermodynamic limit.

\end{document}